\documentclass[aip,jcp,reprint]{revtex4-1}
\pdfoutput=1

\usepackage{hyperref}
\usepackage{amsmath}
\usepackage[usenames,dvipsnames]{color}
\usepackage{graphicx}

\newcommand{\sci}[2]{\ensuremath{#1 \times 10^{#2}}}
\newcommand{\sub}[1]{\ensuremath{_{\textrm{#1}}}}

\begin{document}

\author{Ravishankar Sundararaman}
\affiliation{Department of Materials Science and Engineering, Rensselaer Polytechnic Institute, 110 8th St, Troy, NY 12180 (USA)}

\author{Kathleen Schwarz}\email{kas4@nist.gov}
\affiliation{National Institute of Standards and Technology, Material Measurement Laboratory, 100 Bureau Dr, Gaithersburg, MD, 20899 (USA)}

\title{Evaluating continuum solvation models for the electrode-electrolyte interface: challenges and strategies for improvement}

\begin{abstract}
{\it Ab initio} modeling of electrochemical systems is becoming a key tool for understanding and predicting electrochemical behavior.
Development and careful benchmarking of computational electrochemical methods are essential to ensure their accuracy. 
Here, using charging curves for an electrode in the presence of an inert aqueous electrolyte, we demonstrate 
that most continuum models, which are parameterized and benchmarked for molecules, anions, and cations in solution,
undersolvate metal surfaces, and underestimate the surface charge as a function of applied potential.
We examine features of the electrolyte and interface that are captured by these models,
and identify improvements necessary for realistic electrochemical calculations of metal surfaces.
Finally, we reparameterize popular solvation models using the surface charge of Ag(100) as a function of voltage
to find improved accuracy for metal surfaces without significant change in utility for molecular and ionic solvation.
\end{abstract}

\maketitle

Accurate modeling of the electrode-electrolyte interface is an active challenge for the
computational electrochemical community, and solving it will enable rapid and systematic improvement
of electrocatalysts, reactants, and electrolytes.
Models for the electrode-electrolyte interface must not only describe the 
reactants and surface of interest with high chemical accuracy (using {\it ab initio} techniques such 
as density functional theory (DFT)), but also accurately describe
the changing interfacial charge distribution with potential, to correctly predict
mechanisms and kinetics (onset voltage, barrier heights) of electrochemical reactions.
Calculation of the charge on a given surface at a given voltage requires both the
surface capacitance, and the fluid contribution to the capacitance which is a
thermodynamic average of solvent and electrolyte ion configurations.  
The number of necessary configurations, and the low relative concentration of
electrolyte ions which necessitates inclusion of large numbers of solvent molecules for
meaningful statistics, makes these calculations difficult and computationally expensive.

To make this problem more tractable, a number of approaches have been developed
to perform {\it ab initio} modeling of the electrified interface, from canceling out
the surface charge through a uniform opposite charge across the entire cell,\cite{Neurock}
or through introducing localized classical counter-charges in the effective screening medium approach,\cite{OtaniESM}
to including protons far away from the cell to cancel the charge on the surface.\cite{Norskov}
These approaches do not approximate the spatial distribution of charge in the 
electrolyte, which can significantly influence the structure and energetics of some adsorbates.
An alternative is the solvation model approach with a dielectric continuum description
of the solvent,\cite{PCM-Review} wherein Debye screening due to the electrolyte
cancels out the surface charge.\cite{Letchworth-Weaver2012,Dabo}
These continuum approaches improve upon previous vacuum extrapolation techniques,
and possibly also upon calculations including only one or two explicit water molecule layers,\cite{TwoWaterLayers}
as they include the response of the electrolyte to the charged surface.

Solvation models for use in {\it ab initio} calculations were originally designed 
primarily for the solvation of non-periodic systems of small, neutral molecules.
Fattebert and Gygi\cite{Gygi} modified an isodensity continuum model
(where the spatial extent of the continuum liquid is derived from the electron density)
to make it numerically stable for periodic systems within a plane wave basis,
hence allowing continuum solvation of surfaces.  
Treatment of charged species and interfaces in electrochemical systems
additionally requires the inclusion of the response of ions in the electrolyte.
The simplest approach for this is Poisson-Boltzmann (PB) theory,\cite{PBreview}
which treats the ions as point particles with mean-field interactions.
The drastic approximation of treating the solvent and ions as a continuum
lead to bound charge distributions that are closer to the solute than expected.\cite{Letchworth-Weaver2012}
Fitting the models to solvation energies results in good agreement for
electrostatic interactions between solute and the solvent dielectric
response despite differences in the spatial distributions.
However, the ionic response is not constrained by fits
and tends to be overestimated in the PB approach.

The ionic response plays a smaller role than the solvent dielectric
in the overall energetics of the continuum solvation approach; nonetheless
different approaches have been attempted to alter its spatial extent and intensity.
Jinnouchi and Anderson restricted the ionic response to one solvation shell
away from the solute, creating an approximation of the Stern layer.\cite{jinnouchi,jinnouchi2}
Modifications of the mean-field PB approach to incorporate Stern layer effects
is also an area of active development.\cite{Bazant1, Bazant2,Andelman,Andelman2}
We introduced an alternate approach (NonlinearPCM) based on the classical density-functional theory
of liquids\cite{RigidCDFT} that incorporates a packing limit to stabilize against large ionic response,
but uses the same cavity for the liquid (dielectric) and ionic response.\cite{Gunceler2013}
For high ionic concentrations, an even simpler approach is to use the linearized
Poisson-Boltzmann equation (LinearPCM), which avoids these issues from the beginning by
precluding large build-up of ionic charge.\cite{Letchworth-Weaver2012,Gunceler2013}
Here we focus on such single-cavity models of solvent and electrolyte response in this work, which have
recently attracted attention as a way to perform routine, affordable electrochemical calculations.
These models have been applied to examine electrochemical reactions
including formic acid oxidation\cite{Schwarz2015} and underpotential hydrogen
deposition\cite{Schwarz2016} on platinum, and CO\sub{2} reduction on copper.\cite{Goddard, HeadGordon}
They have also been applied to predict electrochemical capacitance
in new classes of electrodes such as graphene and borophenes for
supercapacitor applications.\cite{GrapheneCapacitance,BoropheneCapacitance}

Careful benchmarking of these models is crucial to understanding their accuracy
for a variety of systems, but so far few comparisons with electrochemical
experiments exist and hence the accuracy of these models is poorly understood.
Prior efforts in this direction include finding the potential at which
specific processes such as changing molecular orientation occur,\cite{Sautet2016}
but the indirect dependence on electrolyte model in these processes
limit the information that can be derived about model accuracy.

A direct test of model accuracy for charging behavior of the electrode,
with readily available corresponding experimental measurements,
is the capacitance of the electrochemical interface.
In particular, the capacitance of the single-crystalline Ag(100) surface is
an ideal test case since it exhibits double-layer behavior without
specific adsorption or surface reconstruction over a wide potential range.
This is seen in at least two experiments in (nearly) non-adsorbing
electrolytes: in aqueous NaClO\sub{4} electrolyte by Hamelin\cite{Hamelin}
and in aqueous KPF\sub{6} electrolyte by Valette.\cite{Valette}
The experimental differential capacitance appears to be nearly symmetric
as a function of potential around the potential of zero charge (PZC) --
more so for KPF\sub{6} than NaClO\sub{4}, suggesting that specific adsorption
of ions (especially anions) is not a significant problem with these electrolytes.
Consequently, here we use these Ag(100) capacitance measurements
to test and refine solvation approaches for electrochemistry.

We perform DFT calculations with various solvation models
using previously established methodology from Ref.~\citenum{CANDLE}.
Briefly, we use the JDFTx code,\cite{JDFTx} with the PBE exchange-correlation functional,\cite{PBE}
the GBRV ultrasoft pseudopotential set\cite{GBRV} at its recommended
plane-wave cutoffs of 20~$E_h$ for orbitals and 100~$E_h$ for charge density.
We utilize a larger fluid spacing of 60.8 \AA~between the metal slabs to capture
the longer range response of the fluid models at lower ionic concentrations (such as 0.01M),
and use truncated Coulomb potentials to minimize interactions between periodic images.\cite{TruncatedEXX}
We include DFT-D2 dispersion corrections\cite{Dispersion-Grimme} for the calculations with
explicit water molecules in DFT to correctly describe the binding energy and distance.
Additionally, because of the increased computational expense of performing the calculations
with explicit water molecules, for these calculations we use a smaller fluid spacing (14.7 \AA),
and limit our calculations to 1 M ionic concentration.

\begin{figure}
\centering
\includegraphics[width=\columnwidth]{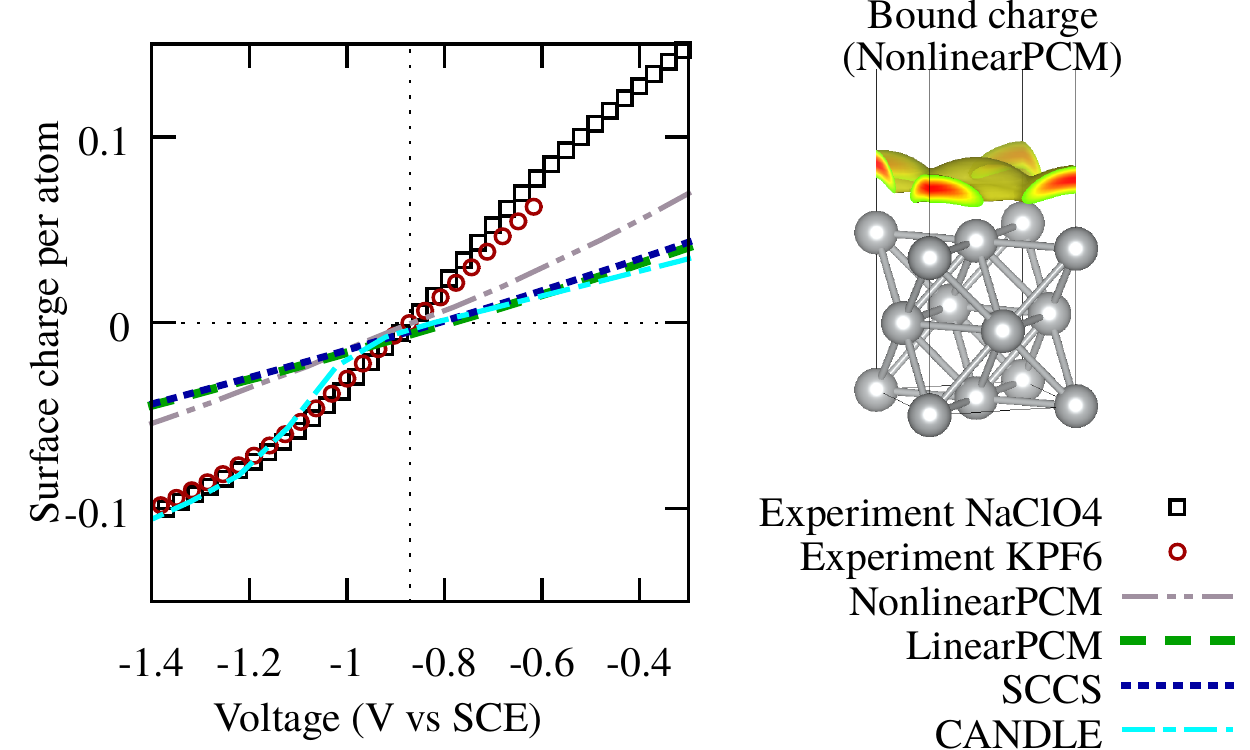}
\caption{Left:  Surface charge as a function of electrode potential at 0.1 M ionic concentration for LinearPCM\cite{Gunceler2013},
CANDLE\cite{CANDLE}, NonlinearPCM\cite{Gunceler2013}, and the SCCS\cite{PCM-SCCS, PCM-SCCS-PB}
solvation models, compared to experimental data from Hamelin~\cite{Hamelin} for NaClO$_4$
and Valette~\cite{Valette} for KPF$_6$.
Right:  Bound charge in the NonlinearPCM continuum solvent at 0.6 V above PZC.
\label{fig:CompPCM}}
\end{figure}

We compare surface charges calculated using different solvation models as a function of potential,
with experimental estimates of the surface charge obtained by digitizing
differential capacitance data from Refs.~\citenum{Hamelin} and \citenum{Valette},
and numerically integrating outwards from the PZC.
Fig.~\ref{fig:CompPCM} shows that most of the commonly used
continuum solvation models in DFT calculations of electrochemical systems
severely underestimate the surface charge of the Ag(100) surface.
In fact, at 0.5 V above the PZC of Ag(100),
the charge on the surface is already underestimated by more than
a factor of two for all of the solvation models, and this
worsens with increasing potential.
We therefore expect that electrochemical reactions that involve transfer
of charge will not be accurately described by any of these models,
especially at potentials far from the PZC.

LinearPCM\cite{Gunceler2013} and SCCS,\cite{PCM-SCCS,PCM-SCCS-PB}
which approximate the electrolyte response
with a linearized Poisson-Boltzmann equation, are the worst performers.
Note that the commonly used VASPsol code\cite{VaspSol} exactly implements the LinearPCM model.
Therefore this underestimation is an issue in the primary solvation models available in all major plane-wave DFT software.
NonlinearPCM, which solves the (full) Poisson-Boltzmann equation, accounting for dielectric saturation
in the solvent response, and nonlinear enhancement and packing effects in the ionic response,\cite{Gunceler2013}
performs marginally better than the linearized models.
All of these models have an ionic response that likely is too close to the solute,
 but this would lead to overestimation rather than underestimation of charge,
and hence is not the reason for the incorrect capacitance.
Instead, they all contain a parameter (or more) to define the spatial extent
of the electrolyte, which is fit to molecular solvation energies,
with no ions or metallic surfaces included in these parameterization datasets.
This procedure of parameterizing solvation models to solely neutral molecules
results in an undersolvation of metals, with charge underestimation as a consequence.
(See Ref.~\citenum{Gunceler2013} for a detailed discussion of the solvation models and their parametrization.)

The CANDLE solvation model also solves the linearized Poisson-Boltzmann equation,
but it is fit to molecules and ions, and adjusts the spatial extent of the electrolyte depending on
the local charge of the solute to describe differences in cationic and anionic solvation.\cite{CANDLE}
Fig.~\ref{fig:CompPCM} shows that the CANDLE capacitance agrees well with experiment for potentials
negative of the PZC, but underestimates it similarly to the other solvation models for positive potentials.
The anionic parameterization, which yields a fluid cavity that is smaller than that for
neutral molecules and cations, appears to be suitable for the metallic surfaces (regardless
of charge), whereas the cationic/neutral parameterization undersolvates the surface.\footnote{The nonlinearity of the CANDLE surface charge curve differs from that reported 
in Ref.~\citenum{CANDLE}, because the previous calculations omitted the contribution to
the electrostatic potential due to the asymmetry correction.}

\begin{figure}
\centering
\includegraphics[width=\columnwidth]{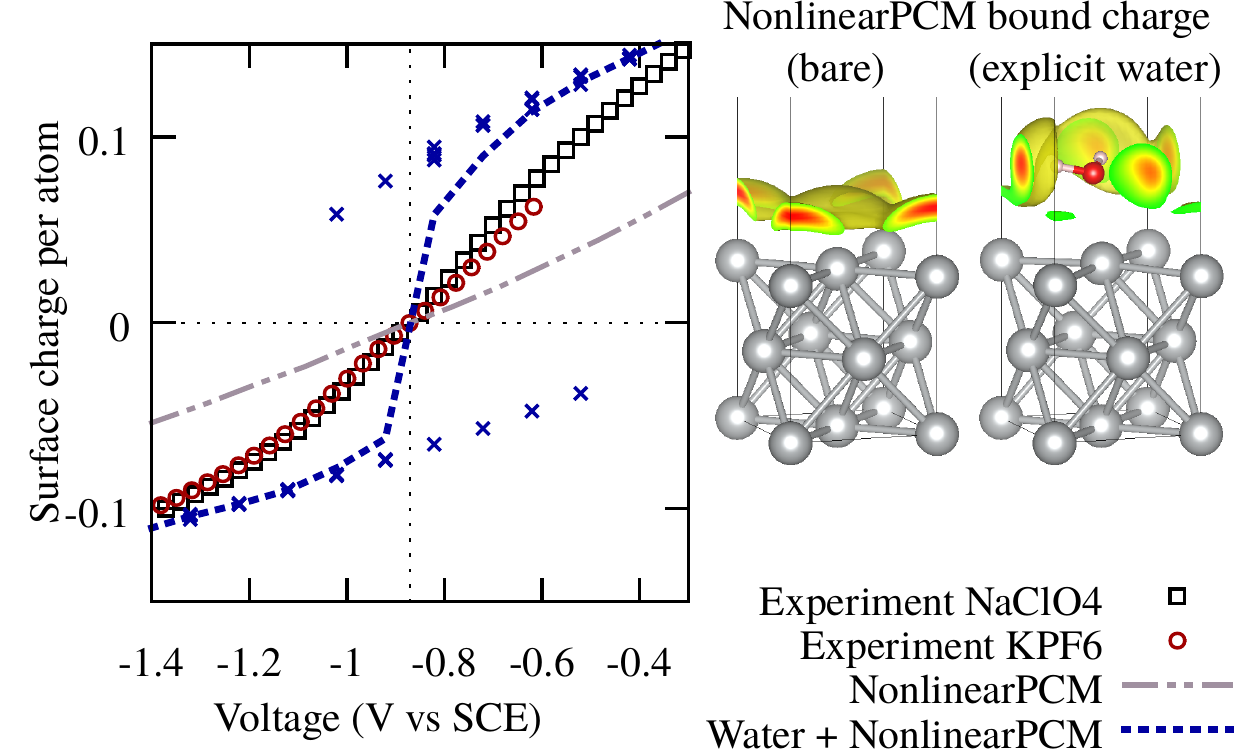}
\caption{Left:  Surface charge as a function of electrode potential for a Ag(100) surface
with continuum solvation, and with a single (DFT) water molecule above the surface.
The blue points are individual minimized configurations of the water,
and the blue line is the thermodynamic average of the two configurations,
using the Gibbs free energies of the configurations at a given potential.
Right:  Bound charge in the continuum solvent above the bare Ag(100)
and surrounding the single explicit water molecule above Ag(100)
at 0.6 V above PZC.
\label{fig:SingleWater}}
\end{figure}

While the bare solvated surface might be an ideal way to test the accuracy of the solvation model,
in practice, DFT calculations frequently include a monolayer or more of explicit (DFT) water molecules.
Fig.~\ref{fig:SingleWater} compares the change in surface charge with voltage between a calculation performed
with one explicit water molecule with nonlinearPCM adjacent and above it, with that of the bare solvated surface.
A larger unit cell with two Ag atoms per surface was used for these calculations
so that explicit water molecules are surrounded by the continuum solvent
and we can avoid geometric issues such as hydrogen bond frustration between
neighbouring water molecules (which would necessitate large unit cells and molecular dynamics).
In this case, there are two free energy local-minimum configurations
of the water molecule, with the water molecule dipole pointing towards
or away from the surface, which differ significantly in surface charge at the same potential.
The global free energy minimum changes from one of these configurations to the other at the PZC,
resulting in a large change in the thermodynamically averaged surface charge
(weighted by Boltzmann factors of the calculated Gibbs free energies).
This average charge therefore exhibits a much steeper slope with potential than experiment
within 0.4 eV of the PZC due to this change in dipole, and a smaller slope comparable to
NonlinearPCM further away from the PZC due to the continuum model response.

The above thermodynamic averaging procedure between two water configurations
grossly approximates features of the surface charging curve that are observed experimentally,
but it is non-trivial to systematically extend this treatment to a larger number of configurations
with multiple water molecules.
With increasing number of molecules, the configurations would be less clearly defined,
and expensive thermodynamic sampling using molecular dynamics will become necessary.
Even if feasible computationally, hybrid treatments combining molecular dynamics
in few explicit layers and continuum solvation above require further method development
to ensure proper matching of the explicit and continuum solvents, while preventing
explicit solvent molecules from drifting into the continuum solvent.

\begin{figure}
\centering
\includegraphics[width=\columnwidth]{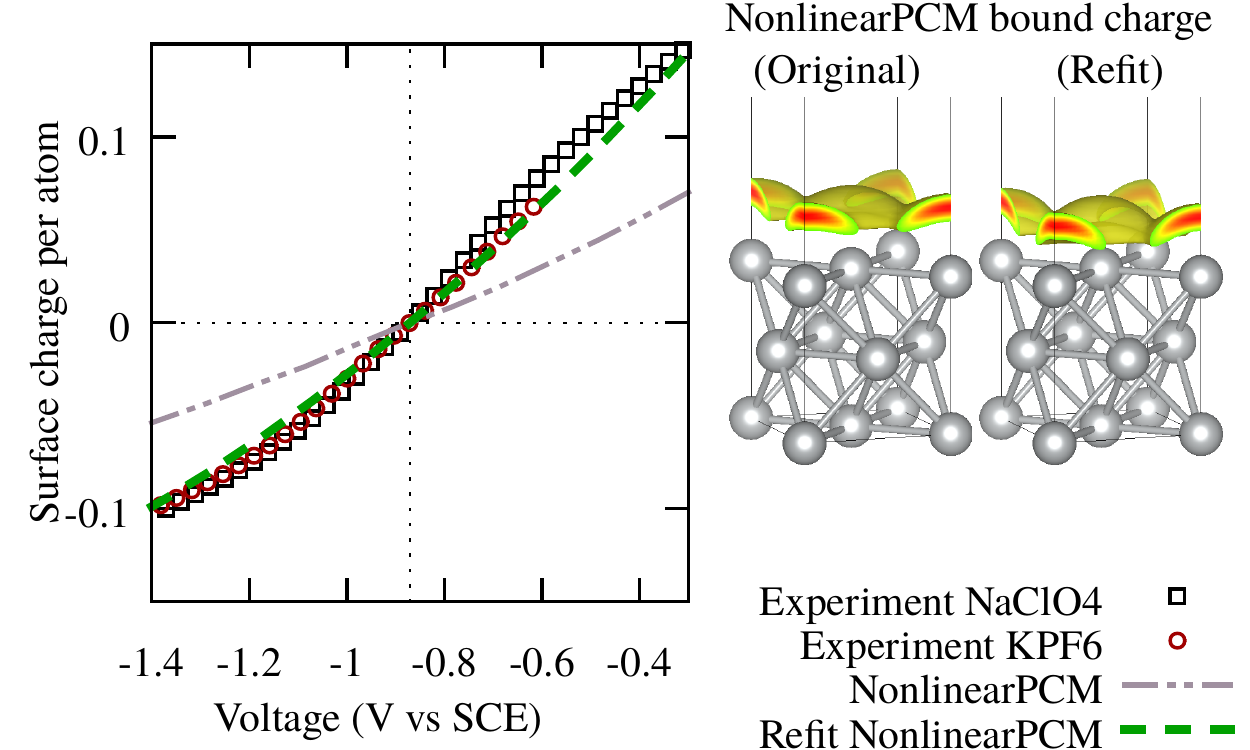}
\caption{Left:  NonlinearPCM with the density cutoff parameter $n_c$ from a molecular fit, Ref.~\citenum{Gunceler2013},
and with a value for $n_c$ that more accurately captures the differential capacitance of Ag(100), for 0.1 M ionic
concentration.  Right:  Bound charge in the fluid for the original NonlinearPCM and the refit NonlinearPCM
at 0.6 V above PZC.
\label{fig:NewNLPCM}}
\end{figure}

Here, we explore an alternate practical approach of reparametrizing solvation models
to potentially correct for the undersolvation of metal surfaces.
In the LinearPCM / VASPsol and NonlinearPCM models, the distance from the solute
at which the continuum solvent appears is controlled by an electron density
threshold parameter $n_c$.\cite{Gunceler2013}
(The SCCS model\cite{PCM-SCCS, PCM-SCCS-PB} uses two electron density parameters $\rho\sub{min}$
and $\rho\sub{max}$ in a formally different but functionally equivalent parameterization.)
To correct the surface charging behavior of Ag(100), we adjust
the value of $n_c$ for the nonlinear model to match the surface charge at the highest potential
reported experimentally in Ref.~\citenum{Hamelin} (0.6~V above PZC).
For the linear model, we report the reparameterization of this model against the surface charge
of the nonlinear model at 1M ionic strength.
Fig.~\ref{fig:NewNLPCM} illustrates that the resulting higher $n_c$ 
(reported in Table~\ref{tbl:params} and discussed further below),
which defines the continuum fluid as beginning closer to the metal leads to
good agreement with the experimental charge/voltage relationship for this surface.

To examine how the reparameterized model performs for other metal surfaces,
we next calculate and compare the differential capacitance of Pt(111).
A single value for the capacitance of the Pt(111) surface has been estimated experimentally,\cite{ExptPt}
and previously reported for continuum and explicit solvation model approaches, as shown in Table~\ref{tbl:Pt111}.
The continuum models predict lower capacitance for the Pt(111) relative to the experimental estimate,
while explicit solvation predicts a higher capacitance, with a similar magnitude of discrepancy.
The refit NonlinearPCM model more closely agrees with the explicit solvation results.
However, the degree to which we accurately predict the capacitance is difficult to determine,
since we are comparing to a single experimental estimate reported with no error bars.

\begin{table}
\centering{\begin{tabular}{ccc}
\hline\hline
Method & Source & Value ($\mu$F/cm$^2$) \\
\hline
Experimental estimate & { Ref.~\citenum{ExptPt}} & 20 \\
\hline
\textbf{Calculated:}\\
Explicit H$^+$ and H$_2$O & Ref.\citenum{Pt111Calc} & 26 \\
LinearPCM & This work & 14 \\
NonlinearPCM & This work & 17 \\
CANDLE & This work & 11 \\
Refit NonlinearPCM & {\bf This work} &  29 \\
Refit LinearPCM & {\bf This work} &  31 \\
\hline\hline
\end{tabular}}
\caption{Capacitance for Pt(111) by different methods, including the reparameterization for the last two models described here.}
\label{tbl:Pt111}
\end{table}

\begin{figure}
\centering
\includegraphics[width=\columnwidth]{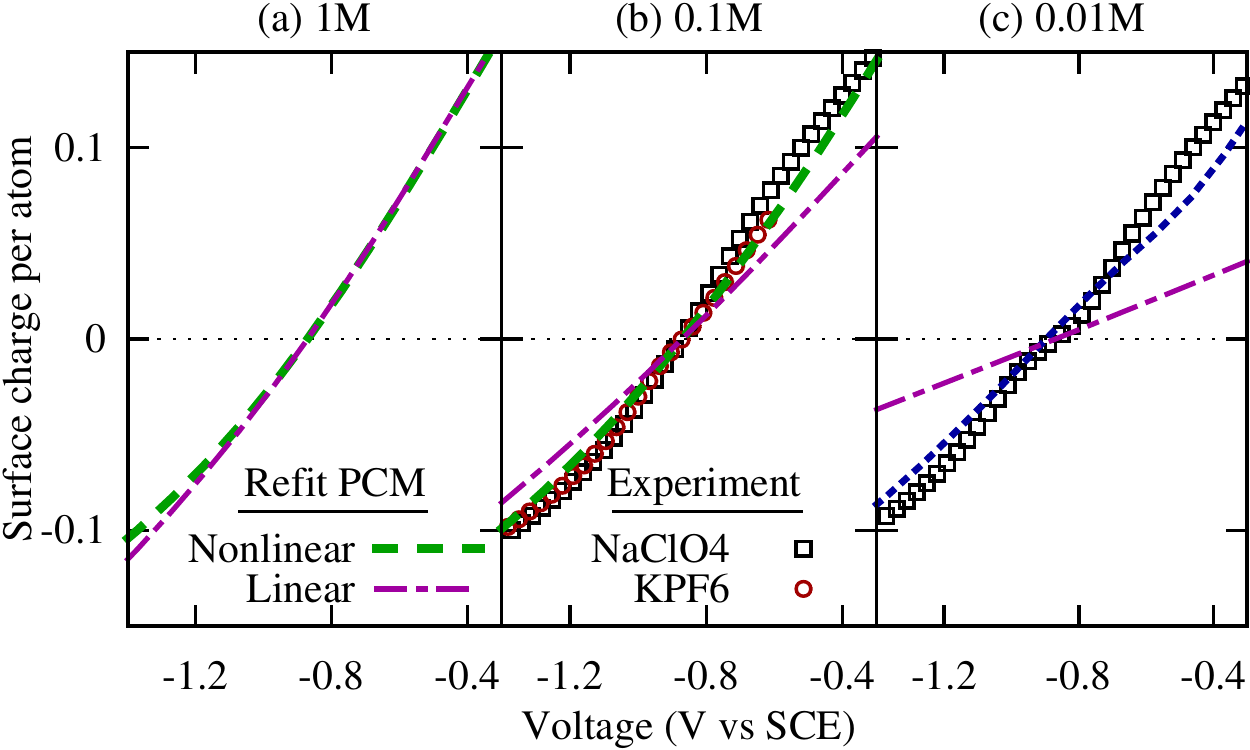}
\caption{Comparison of surface charge predicted by refit NonlinearPCM,
and refit LinearPCM with experimental data from Hamelin~\cite{Hamelin}
for NaClO$_4$ and Valette for KPF$_6$~\cite{Valette}
for different ionic concentrations: (a) 1M, (b) 0.1M and (c) 0.01M.
NonlinearPCM qualitatively captures the variations in capacitance
with potential and ionic concentration, whereas LinearPCM fails.
\label{fig:ConcDep}}
\end{figure}

To understand the transferability of our refit models for use under other common electrochemical
conditions, we next evaluate the behavior of the refit models as a function of ionic concentration.
Figure~\ref{fig:ConcDep} illustrates the success of the nonlinear model in capturing the 
surface charge characteristics as a function of decreasing ionic strength,
and the drastic failures of the linear model.  This is in agreement with
previous qualitative explorations of the importance of nonlinear behavior~\cite{Dabo}.
We note limitations in the nonlinear model, however.
Solely reparameterizing the solvation model does not capture all of the nonlinearity
of the surface charging curve with voltage, and in particular underestimates
the Gouy-Chapmann behavior seen experimentally at lower ionic strengths at
voltages near the PZC, which requires further investigation.

While the refit models are promising for metal surfaces, the reparameterization 
of only the cavity threshold $n_c$ results will necessarily oversolvate
molecules and ions for which the original $n_c$ was optimized.
This can be mitigated somewhat by refitting the second parameter
in these models which capture the non-electrostatic contribution
to solvation, represented by a cavity surface tension parameter $t$.
(See Ref.~\citenum{Gunceler2013} for more details about the fit parameters.)
Table~\ref{tbl:refit} lists the mean absolute errors for the solvation energies of a set
of molecular and ionic species for the original and refit models.
The reparameterized nonlinearPCM performs better for cations,
slightly worse for anions and neutral molecules,
and overall marginally better for the combined set.
The resulting improvements for metal surfaces though,
makes this parameterization preferred over that of the original.
Table~\ref{tbl:params} lists the original and refit parameters,
so that these new values can be utilized for future calculations.

\begin{table}
\centering{\begin{tabular}{rcccc}
\hline\hline
Model              & Neutrals &  Cations  &  Anions & Combined \\
\hline
NonlinearPCM       &    1.28   &  16.08    &  27.03   &   7.55 \\
Refit NonlinearPCM &    1.44   &  14.40    &  27.61   &   7.51 \\
LinearPCM          &    1.27   &  2.10     &  15.09   &   3.59 \\
Refit LinearPCM    &    2.57   &  19.73    &  12.50   &   6.68 \\
\hline\hline
\end{tabular}}
\caption{Mean absolute errors in kcal/mol (1 kcal/mol is 0.0434 eV) for NonlinearPCM and LinearPCM with
original parameterization, and with refit $n_c$ and surface tension $t$,
for the set of molecules, anions, and cations in Ref.~\citenum{CANDLE}.
\label{tbl:refit}}
\end{table}

\begin{table}
\centering{\begin{tabular}{rccc}
\hline\hline
Parameters & $n_c\ [a_0^{-3}]$ & $t\ [E_h/a_0^2]$ & V\sub{SHE} [V] \\
\hline
NonlinearPCM       & \sci{1.0}{-3} & \sci{9.5}{-6} & 4.62$\pm$0.09\cite{Gunceler2013} \\
Refit NonlinearPCM & \sci{2.2}{-3} & \sci{2.6}{-5} & 4.33$\pm$0.09 \\
LinearPCM          & \sci{3.7}{-4} & \sci{5.4}{-6} & 4.68$\pm$0.09\cite{Gunceler2013} \\
Refit LinearPCM    & \sci{1.4}{-3} & \sci{4.2}{-5} & 4.10$\pm$0.05 \\
\hline\hline
\end{tabular}}
\caption{Parameters for the original and refit NonlinearPCM and LinearPCM models,
and the corresponding reference electrode potential $V\sub{SHE}$ calibrated to
experimental potentials of zero charge of single-crystalline Ag, Au and Cu surfaces.}
\label{tbl:params}
\end{table}

Table~\ref{tbl:refit} shows that the reparametrization worsens the accuracy of LinearPCM
for molecules and cations, while slightly improving it for anions,
again a consequence of anions being undersolvated to start with.
Table~\ref{tbl:params} reports the corresponding parameters
which can be used for improving the predictions for metal surfaces in
implementations of NonlinearPCM and LinearPCM, such as in VASPsol\cite{VaspSol} and JDFTx.\cite{JDFTx}

This work illustrates that design of continuum solvation models for describing charged metal electrodes
must include short-range cavity parameterization that simultaneously leads to correct molecular
and ionic solvation energies, as well as differential capacitance curves for metal surfaces.
Describing the nonlinearity in the ionic response is further important for describing
these systems correctly at lower ionic strengths.
Benchmarking against the surface charging for Ag(100) provides a way to compare performance of electrolyte models.
Our reparameterization of both the linear and nonlinear solvation models provide better options for 
electrochemical calculations in which the charge of the surface might be important.
Despite the fact that we have demonstrated the serious limitations of the LinearPCM model
for lower ionic strengths, the linear model is currently more widely available (in codes other than JDFTx, for example).
Therefore, we recommend using the reparameterized NonlinearPCM model for computational electrochemical calculations of metallic surfaces,
but if that is not possible, we suggest utilizing the reparameterized LinearPCM model at 1 M ionic strength as a work around,
regardless of the expected or actual experimental ionic strength. 

Additionally, we note that further careful experimental measurements of the differential capacitance of several single-crystalline
metal surfaces in non-adsorbing electrolytes are necessary to guide the development of the
next generation of solvation models that can achieve simultaneous accuracy for molecules, ions and electrodes.

We acknowledge helpful conversations and feedback from Thomas Moffat, Kendra Letchworth-Weaver,
and  Erdal Uzunlar.

\bibliographystyle{apsrev}
\makeatletter{}

\end{document}